%% file: Main.tex
\documentclass{article}
\usepackage{spconf, graphicx, amsmath, hyperref,setspace,enumitem, amssymb}
\usepackage{subcaption}
\usepackage{multirow}
\usepackage[dvipsnames]{xcolor}
\usepackage{mathtools}

\newcommand\tabref{Table~\ref}

\input{def.set}

\title{Neural Feature Predictor and Discriminative Residual Coding\\for Low-Bitrate Speech Coding}

\name{Haici Yang$^1$, Wootaek Lim$^2$, Minje Kim$^1$\thanks{This work was supported by Electronics and Telecommunications Research Institute (ETRI) grant funded by the Korean government
(22ZH1200; ``The research of the basic media contents technologies").}}
\address{ $^1$Indiana University, Luddy School of Informatics, Computing, and Engineering, Bloomington, IN, USA\\
$^2$ Electronics and Telecommunications Research Institute, Daejeon 34129, South Korea}

\ninept
\begin{document}

\maketitle

\begin{abstract} 

Low and ultra-low-bitrate neural speech coding achieves unprecedented coding gain by generating speech signals from compact speech features. This paper introduces additional coding efficiency in neural speech coding by reducing the temporal redundancy existing in the frame-level feature sequence via a recurrent neural predictor. The prediction can achieve a low-entropy residual representation, which we discriminatively code based on their contribution to the signal reconstruction. The harmonization of feature prediction and discriminative coding results in a dynamic bit allocation algorithm that spends more bits on unpredictable but rare events. As a result, we develop a scalable, lightweight, low-latency, and low-bitrate neural speech coding system. We demonstrate the advantage of the proposed methods using the LPCNet as a neural vocoder. While the proposed method guarantees causality in its prediction,
the subjective tests and feature space analysis show that our model achieves superior coding efficiency compared to LPCNet and Lyra V2 in the very low bitrates.

\end{abstract}



\input{introduction}

\input{model}

\input{bitrate}

\input{experiments}
\input{results}

\input{conclusion}
\bibliographystyle{IEEEtran}
\bibliography{mjkim}

\end{document}

%% file: introduction.tex
\section{Introduction}

A speech codec, in general, comprises modules for speech compression, quantization, and reconstruction. It has been used in various communication and entertainment applications after standardization \cite{BessetteB2002amrwb, SchroederM1985celp} or open-sourcing \cite{ValinJM2012opus}. The common goal in speech coding is to achieve the maximum coding gain, i.e., maintaining the perceptual quality and intelligibility of the reconstructed speech signals with a minimum bitrate.

The involvement of neural networks has greatly benefited the coding trade-off, effectively eliminating the codes’ redundancy while improving the reconstruction quality. More recently, the advances of generative models and their applications in speech coding led to a trend in very low-bitrate speech codecs. The first WaveNet-based speech codec \cite{KleijnW2018wavenet} demonstrates the usage of neural synthesis in both waveform and parametric coding. The latter is more favored in subsequent studies because of its inherent advantages in dealing with very condensed speech features. These neural vocoders work on the decoder side, leveraging the powerful neural synthesis architecture. Their encoding parts are relatively simplified, relying on existing Codec 2 codes  \cite{codec2} as in the original WaveNet-based speech codec \cite{KleijnW2018wavenet} or the dimension-reduced frequency-domain speech representations, e.g., cepstrum features \cite{ValinJ2019lpcnet, ValinJ2019lpcnetcoding}, and LPC analysis \cite{KlejsaJ2019samplernn}. 

In this line of work, the performance bottleneck comes from the very compact codes, leading to poorer reconstruction quality. To mitigate the issue, some efforts apply more complex models in the encoder to improve the quality of the features \cite{kim2021neurally, yoshimura2018wavenet} or use generative models for post-processing \cite{zhao2018convolutional, skoglund2019improving, biswas2020audio} at the end of the existing codec to facilitate signal restoration. However, the output performance is still bounded by the quality of the coding features. 

End-to-end neural codecs that train the encoder, quantizer, and decoder jointly work as an alternative to the low-bitrate generative speech vocoders \cite{KankanahalliS2018icassp, ZhenK2019interspeech, ZhenK2022taslp, Zeghidour2021soundstream}.
In this way, the neural encoder participates in removing the redundancy in the source signal and produces features that are more associated with the decoder, in contrast to the traditional speech features. Various other methods have been developed to improve the quality of the features, regarding robustness \cite{casebeer2021enhancing, Lim2020robust}, scalability \cite{jiang2022cross} and the variability issues \cite{kleijn2021generative}.

\begin{figure}
    \centering
    \includegraphics[width=.8\columnwidth]{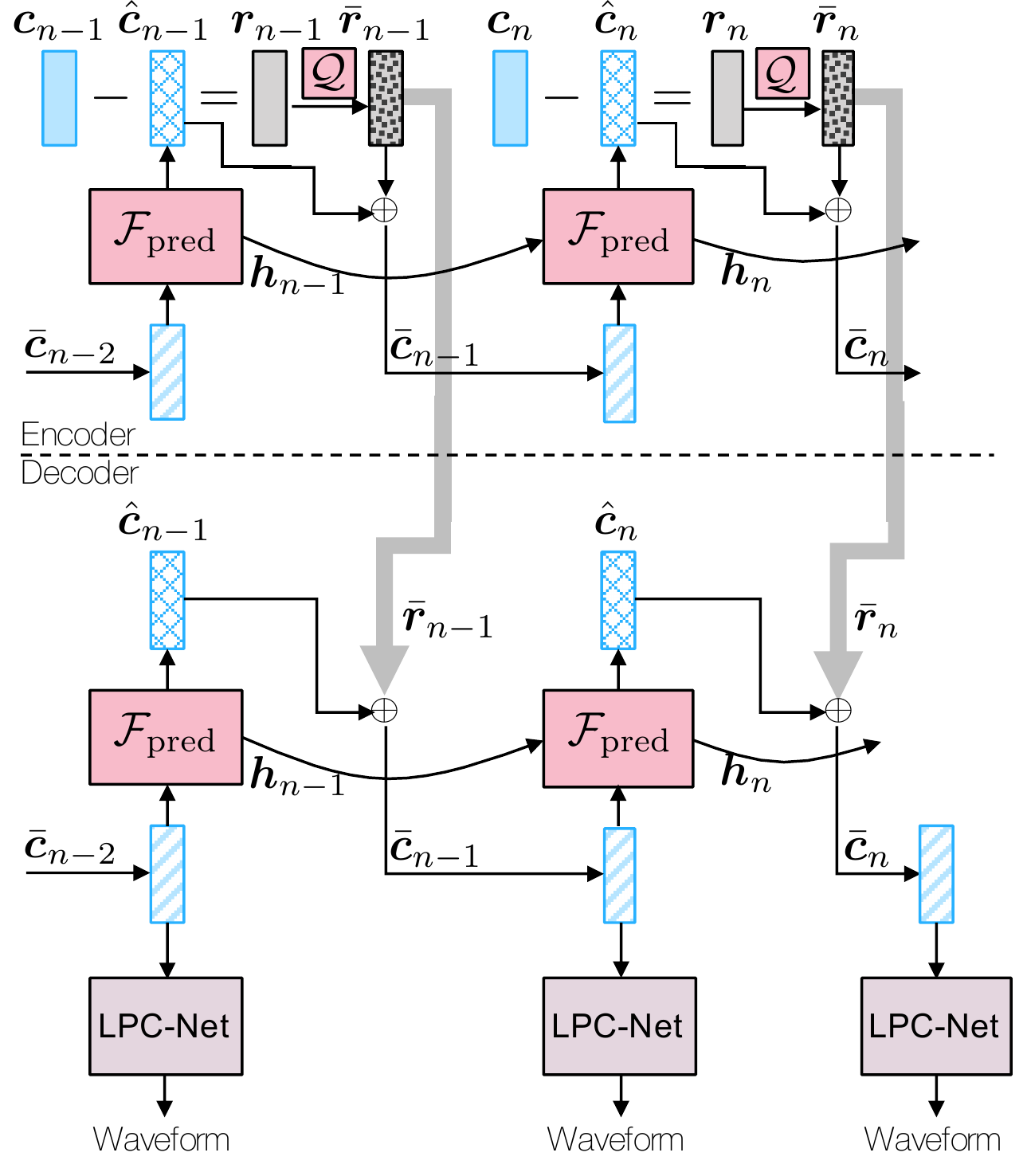}
    \caption{The overview of the proposed neural speech coding system with feature predictors and LPCNet-based vocoder.}
    \label{fig:overview}
    \vspace{-6mm}
\end{figure}

However, end-to-end codecs tend to suffer in very low-bitrates cases ($<$2 kbps) because that requires the coding features to be extremely small and expressive simultaneously. To deal with that, an ultra-low bitrate codec \cite{SiahkoohiA2022interspeech} borrows the embeddings from a self-supervised training task to increase the expressiveness of the state-of-the-art codec SoundStream's features \cite{Zeghidour2021soundstream}, and can obtain a decent speech quality with a very low bitrate, $0.6$ kbps.
TF-Codec \cite{jiang2022predictive} addresses the problem by reducing the temporal redundancy in the latent features with a predictive model and reports decent reconstruction quality at $1$ kbps. However, both models entail high complexity. Besides, because TF-Codec's prediction model runs on a latent space that requires a specific pair of encoder and decoder, it brings an extra cost for other existing codecs to mount its predictive module directly.

In this paper, we aim at a low-bitrate, low-delay, and low-complexity neural speech codec that utilizes neural feature prediction to reduce the temporal redundancy from the sequence of feature frames.
We introduce a gated recurrent unit (GRU)-based \cite{ChoK2014emnlp} frame-level feature predictor that can forecast the feature vector at time frame $t$ using its preceding frames. Since the decoder also employs the exact feature predictor, it can ``generate" most of the feature vector at no cost of bitrate, while the imperfectly generated feature vectors are compensated by the coded residual coming from the encoder side. 
Additionally, we employ \textit{discriminative coding} in the residual space. This idea is demonstrated in source-aware neural audio coding by distinguishing speech and noise sources in the latent feature space \cite{YangH2021sanac}. In this paper, we use different entropy coding strategies at each frame depending on the amount of information they carry. Compared to the TF-Codec, our model explicitly codes only the prediction residuals, and the proposed predictive modules are designed to work in combination with existing low-complexity neural codecs. In particular, we are based on the efficient LPCNet-based speech coding framework \cite{ValinJ2019lpcnetcoding}, and our analysis and model training mainly focuses on the cepstral coefficients.

%% file: model.tex
\section{The proposed predictive coding}
\subsection{Overview}
In conjunction with the LPCNet's sample-level vocoding, the proposed feature prediction model performs hierarchical prediction: first in the feature space and then in the sample level. As shown in Fig \ref{fig:overview}, the frame-level feature predictor $\mathcal{F}_{pred}$ works on both the encoder and decoder sides. The encoder computes and quantizes the frame-level prediction residuals $\bar{\bm r}$ and passes them to the decoder. Then, the decoder adds the received residuals to its own feature predictions $\hat{\bm c}$ to obtain the recovered frame-level features. The sample-level predictive coding (i.e., the LPCNet vocoder) works only on the decoder that synthesizes waveform samples $\hat{\bm s}$ from the recovered features as the codec's output.


\subsection{The frame-level feature predicion}

\subsubsection{Feature predictor}
We apply a WaveRNN-based model \cite{KalchbrennerN2018wavernn} to make a frame-level prediction on the 18-dimensional continuous cepstral coefficients. 
WaveRNN explicitly considers the output at time $n-1$ as the estimation of the $n$-th's sample. 
In our frame-by-frame feature prediction scenario, the recurrent neural network  $\mathcal{H}(\cdot)$ takes in previous hidden state $\bm h_{n-1}$ and the previous feature vector $\bm c_{n-1}$, to predict the next frame $\hat{\bm c}_n $. Additionally, we condition the frame-level prediction with pitch parameters $\bm m_n$ (period and correlation) used in LPCNet. Our model consists of two gated recurrent unit (GRU) layers \cite{ChoK2014emnlp}, with 384 and 128 hidden units, respectively, followed by a fully connected layer. The feature predictor $\mathcal{F}_\text{pred}: \bm c_{n-1} \mapsto \hat{\bm c}_n$ can therefore be recursively defined as,
\begin{equation}\label{eq:prediction}
    \bm h_n = \mathcal{H}(\bm c_{n-1}, \bm h_{n-1}, \bm m_n), \quad
    \hat{\bm c}_n = \text{tanh}(\bm W \bm h_{n}), 
\end{equation}
where $n$ represents the time-domain index. We found the results are more stable by scaling input and output features to the range of $[-1,1]$. To this end, the output linear layer employs a tanh activation after a linear combination with parameter $\bm W$. Biases are omitted for brevity.

We optimize the model by minimizing the mean squared error (MSE) between the prediction and target $\mathcal{L} = MSE(\bm c_n, \hat{\bm c}_n)$. We chose it over the maximum log-likelihood approach with explicit Gaussian modeling of the features because modeling the cepstrum coefficients with Gaussian distributions was unreliable.

\subsubsection{Feature residual coding}


We employ the predictor in both the encoder and decoder to cover the information that can be inferred from the temporal dependency. Thus, for the decoder to recover the features, it is only necessary to provide the decoder with the residuals between the prediction and ground-truth features. This kind of explicit residual coding can lead to a more efficient coding scheme, given that our predictor model makes reliable predictions, especially in the areas of smooth signals, reducing the entropy of the residual. 


The primary pipeline for residual coding is then summarized recursively as follows:
\begin{align}
    \text{Encoder:} \quad \hat{\bm c}_n &=\mathcal{F}_\text{pred}(\bar{\bm{c}}_{n-1}) \label{enc_pred}\\
    \bm r_n &= \bm c_n - \hat{\bm c}_n\\
    \bar{\bm r}_n &= \mathcal{Q}(\bm r_n)\quad \text{(send it to the decoder)} \label{qtz}\\
    \bar{\bm c}_n &= \hat{\bm c}_n + \bar{\bm r}_n \quad \text{(input for the next round } n+1 \text{)} \label{add-back}
\end{align}

The encoder explicitly computes the residual $\bm r_n$, and then the quantizer $\mathcal{Q}(\cdot)$ converts it into a bitstring $\bar{\bm{r}}_n$ as the final code.
Note that we opt to input the \textit{noisy feature} $\bar{\bm c}$ instead of the original feature $\bm c$ into the encoder's feature predictor (eq. \eqref{enc_pred}) in order to match the encoder's output to the decoder's circumstance. Since the decoder does not have access to the original features, it has no choice but to use the noisy ones $\bar{\bm c}$ as the predictor's input. Therefore, by repeating the decoder's behavior in the encoder, we aim to guarantee that the residuals provided by the encoder are the accurate compensation for the decoder's feature prediction.  

The decoder first pre-computes the prediction $\hat{\bm c}_n$, and then supplement it with the quantized residual $\bar{\bm r}_n$ received from the encoder to finalize the feature reconstruction $\bar{\bm c}_n$.
\begin{align}
    \text{Decoder:} \quad \hat{\bm c}_n &=\mathcal{F}_\text{pred}(\bar{\bm{c}}_{n-1})\\
    \bar{\bm c}_n &= \hat{\bm c}_n + \bar{\bm r}_n.
\end{align}
When running $\mathcal{F}_\text{pred}$ in either the encoder or decoder, we starts with zero-initialized input $\bar{\bm c}_0 = \bm 0$ , and iteratively update the input tensor with the model predictions . 

\subsection{The sample-level vocoder: LPCNet}

We borrow LPCNet to complete time-domain synthesis on the decoder side. LPCNet takes as input pitch parameters $\bm m_n$ and cepstrum features $\bm c_n$. Then, it integrates LPC analysis into the neural generative model, i.e., $p_t = \sum_{\tau=1}^T a_\tau \hat{s}_{t-\tau}$, which computes the prediction $p_t$ for the sample index $t$ by using $T$ previously estimated samples $\hat{s}_{t-T:t-1}$. In this way, the burden of spectral envelop modeling is taken away from the neural network. The prediction coefficient $a_\tau$ is computed only from the 18-band Bark-frequency cepstrum (the transmitted code in the original LPCNet coder \cite{ValinJ2019lpcnetcoding}), forming a very compact bitstring. 
On top of the DSP-based linear prediction, LPCNet also employs a WaveRNN network $\mathcal{G}$ to estimate the prediction residuals ${e}_t$ directly in a causal manner:
\begin{equation}
    \hat{e}_t = \mathcal{G}(p_t, \hat{s}_{<t}, \hat{e}_{<t}), \quad
    \hat{s}_t = p_t + \hat{e}_t, 
\end{equation}
where the quality of the estimated excitation signal $\hat{e}_t$ significantly matters for better speech quality. The network $\mathcal{G}$ mainly consists of two GRU layers, followed by two fully connected layers.

We employ the same LPCNet vocoder for our coding system, except that its input feature $\bm c_n$ is replaced with our proposed feature reconstruction $\bar{\bm{c}}_n$, necessitating re-training the vocoder. Note that in the speech coding version \cite{ValinJ2019lpcnetcoding}, LPCNet does take its own compressed feature representation as input, whose compression ratio is something our method competes against.

%% file: bitrate.tex
\section{The proposed discriminative residual coding}

\begin{figure}[t]
    \centering
    \includegraphics[height=0.85in]{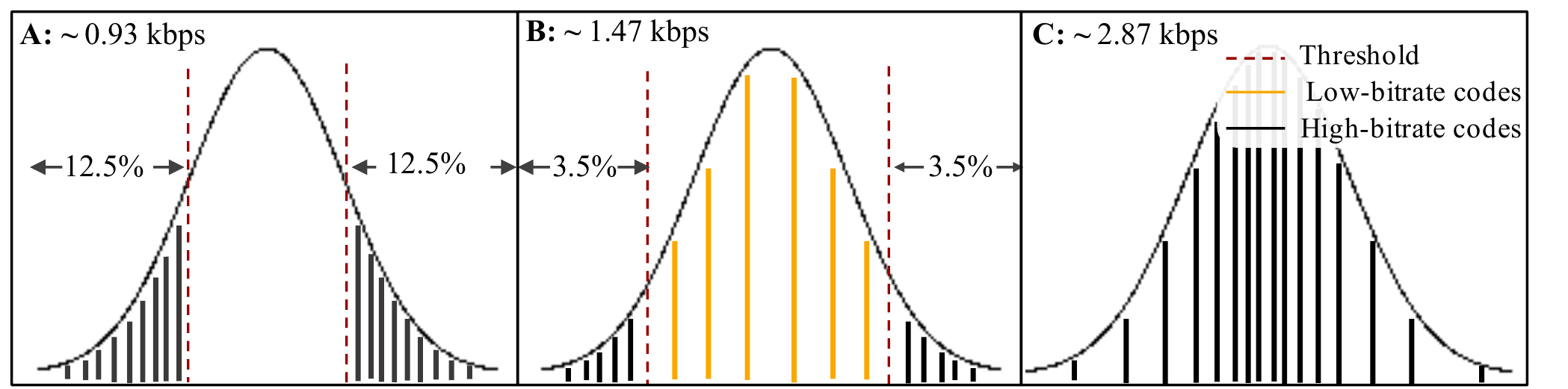}
    \caption{The proposed thresholding mechanism uses different quantization schemes depending on the target bitrate.}
    \label{fig:bitrate}
     \vspace{-4mm}
\end{figure}


To improve the coding gain further, we apply \textit{discriminative} coding to the residual signals $\bar{\bm r}_n$, which is the information sent to the receiver in place of the full cepstrum $\bm{c}_n$. 
Due to the overall smoothness of speech, the prediction in the cepstrum domain results in the residuals that follow a Gaussian distribution with zero mean and small variance. The larger residual values mainly occur in transient events, such as plosives.
To fully make use of the residual signal's statistical advantage, we apply \textit{discriminative} coding that distinguishes the more ``code-worthy" frames from the rest by thresholding the $L_1$ norm of the residuals. This way, frames with significant residual energy are assigned more bits, and bits assigned to the less significant frames are minimized. 



Scanning the entire training set, we define a threshold value $\theta$ depending on the target bitrate. The quantization process eq.\eqref{qtz} is therefore expanded to: 
\begin{equation} 
  \bar{\bm r} = \mathcal{Q}({\bm r}) =  
    \begin{cases}
      \mathcal{Q}_\text{L}(\bm{r}) & \text{if } ||\bm r||_1 \geq \theta \\
      \mathcal{Q}_\text{S}(\bm{r}) & \text{otherwise},
    \end{cases}       
\end{equation}
with $\mathcal{Q}_\text{L}$ and $\mathcal{Q}_\text{S}$ representing quantization schemes with large and small $L_1$ norms that use large and small codebooks, respectively. Particularly, when the target bitrate is extremely low, we \textit{discard} low $L_1$ norm frames entirely, i.e., $\mathcal{Q}_\text{S}(\bm{r})=\bm{0}$. 

The thresholding mechanism is illustrated in Fig.\ref{fig:bitrate}. The low-bitrate scheme ($\sim$ 0.93 kbps) uses $\mathcal{Q}_\text{L}$ for the top 25 \%  residual frames while discarding the rest without any coding. The intermediate bitrate ($\sim$ 1.47 kbps) case keeps the top 7\% for the $\mathcal{Q}_\text{L}$ quantization and the rest 90\% for $\mathcal{Q}_\text{S}$. The $\sim$ 2.87 kbps case uses $\mathcal{Q}_\text{L}$ quantization for all residual frames with no thresholding. 

Similar to LPCNet's coding scheme, we separately code the first component $\bm r_0$ of the residuals vector and the rest of dimensions $\bm r_{1-17}$; Note here that we dropped the frame index $n$ and used subscript to indicate one of the 18 cepstrum coefficients within the frame. Also, since we noticed that $\bm r_0$ and $\bm r_{1-17}$ have different $L_1$ norm distributions, we define thresholds and apply discriminative coding to the scalar and vector components independently.

\begin{table}[]
    \centering
    \resizebox{\columnwidth}{!}{%
    \begin{tabular}{c||c|c||c|c||c|c}
        Target bitrate (kbps) & \multicolumn{2}{c||}{$\sim$ 0.93} & \multicolumn{2}{c||}{$\sim$ 1.47} & \multicolumn{2}{c}{ $\sim$ 2.87} \\
        \hline
        $\mathcal{Q}_\text{L}$ percentage & \multicolumn{2}{c||}{25 $\%$} & \multicolumn{2}{c||}{7 $\%$} & \multicolumn{2}{c}{ 100 $\%$} \\
        \hline
      \multicolumn{7}{c}{Codebook Size (bits) : Bits-per-frame according to Huffman coding}\\
        \hline
        Stages & 1st & 2nd & 1st & 2nd & 1st & 2nd\\
        \hline
        $\mathcal{Q}_\text{L}(\bm{r}_0)$ & 8 : 7.0 & - & 8 : 7.4 & - &8 : 7.2 & - \\
        \hline
        $\mathcal{Q}_\text{S}(\bm{r}_0)$ & - & - & 4 : 2.9 & - & - & - \\
        \hline
        $\mathcal{Q}_\text{L}(\bm{r}_{1:17})$ & 10 : 9.8 & 10 : 9.9 & 10 : 9.2 & 10 : 9.4 & 10 : 9.2 & 10 : 9.6\\
        \hline
        $\mathcal{Q}_\text{S}(\bm{r}_{1:17})$ &  - & - & 9 : 8.0 & - & - & -\\
    \end{tabular}
    }
    \caption{Codebook sizes and bitrate assignments.}
    \label{tab:codebooks}
    \vspace{-3mm}
\end{table}

Table \ref{tab:codebooks} summarizes how we conduct discriminative and multi-stage quantization depending on the target bitrate. For scalar quantization, we use the same codebook size of $2^9=512$ in all $\mathcal{Q}_\text{L}$ cases, while only $16$ codewords for $\mathcal{Q}_\text{S}$ in the mid-bitrate case or skips coding in the low-bitrate case. All scalar quantizers use a single-stage quantization scheme. 
As for the VQ for $\bm{c}_{1:17}$, we employ either one or two-stage quantization for $\mathcal{Q}_\text{L}$ with a codebook of size 1024 in each stage; $\mathcal{Q}_\text{S}$ cases use a single 512-size codebook or skip coding in the low-bitrate case ($\sim$0.93). We also estimate the bitrate considering Huffman coding by computing the frequencies $\bm p$ of all codewords from coding randomly-selected 2-second segments per training samples and derive the average bit-per-frame by $\sum_i^N \bm p_ilog_2\bm p_i$. 
Apart from the bits we have stated in the \tabref{tab:codebooks}, we also need to count in the bits for coding pitch parameters in LPCNet's original way, which takes up 0.275 kbps. We use the bitrates of Huffman coding in the rest of the paper, although it is close to the bitrates based on the codebook. In the $\sim$0.93kbps case, for example, the target bitrate in our table is calculated by $0.25 \times(7\times100+(9.9+9.8 )\times100) + 275 = 932 \text{bps}$, given that each frame is for 10 ms (meaning 100 frames per second), and only $25\%$ of the frames are coded in this example.



%% file: experiments.tex
\section{Experiments}
\subsection{Data}
We use the Librispeech \cite{PanayotovV2015Librispeech} corpus's \texttt{train-clean-100} fold for training, and \texttt{dev-clean} for validation, at 16kHz sampling rate. 
18 Bark-scale cepstral coefficients are produced for each 20 ms frame with an overlap of 10ms. In addition, we extract and quantize the 2-dimensional pitch parameters using LPCNet's open-sourced framework. 

\subsection{Training}
The training process consists of three steps and is conducted sequentially: prediction model training, codebook learning, and vocoder training. Hence, the results from the preceding steps will be used in the following training. 
Compared to a potential end-to-end learning approach, our modularized learning can circumvent the issue of dealing with non-differentiable quantization. 

Both the feature predictor and the vocoder will eventually operate in a synthesis mode, where the inputs to the model are the synthesized results from the previous step. Therefore, we add noise to the input during training for a more robust development, as suggested in  \cite{ValinJ2019lpcnetcoding,JinZ2018fftnet}. 
Finally, the vocoder is finetuned with the quantized input features.

Codebook training is based on the residuals ${\bm{r}}$ produced from the encoder of the feature predictor $\mathcal{F}_{pred}$. For both vector and scalar codebooks, we run k-means clustering and pick the learned centroids as the codewords. 
When generating residuals for codebook training, the encoder skips the quantization step (eq. \eqref{qtz}) but will consider the residual thresholding. That is, the residual $\bm r_n$ will be added back to the prediction result $\hat{\bm c}_n$ (as in eq. \ref{add-back}) only if $||\bm r||_1 \geq \theta$. 
We randomly pick 2-second segments from each utterance in training set to generate the residual vector for codebook training. Codebooks are trained exclusively for each bitrate.


The feature predictor model we used in the experiments contains  $0.65$M parameters, and the entire codec, including the LPCNet vocoder, has $2.5$M parameters. Our codec is suitable for the real-time coding task because of the causality preserved in the frame-level prediction. The algorithmic delay of our codec is $75$ ms, to which the LPCNet vocoder contributes $60$ms-latency from its convolution operation. Another $15$ ms-delay comes from our feature predictor, which occurs while waiting for the ground-truth cepstral-frame of $10$ms with an extra $5$ms look-ahead to compute a cepstrum.

\subsection{Evaluation and baseline}

We employ two state-of-the-art low-bitrate codecs as baselines, LPCNet at 1.6kbps and Lyra V2 \footnote{https://opensource.googleblog.com/lyra-v2-a-better-faster-and-more-versatile-speech-codec.html} at 3.2kbps. Lyra V2 is an improved version of Lyra\footnote{https://ai.googleblog.com/lyra-new-very-low-bitrate-codec-for.html} \cite{kleijn2021generative}, integratin SoundStream \cite{Zeghidour2021soundstream} in its original architectures for a better coding gain.


We perform a MUSHRA test \cite{mushra} on our codec at three different bitrates and the two baselines. Ten gender-balanced clean utterances from the LibriSpeech \texttt{test-clean} set are used. The trials also include a hidden reference and a low-pass-filtered anchor at 3.5kHz. Ten speech experts participated in the test, and no one was excluded per the listener exclusion policy.

%% file: results.tex
\section{Results}


\begin{figure}[t]
        \centering
        \includegraphics[height=1.5in]{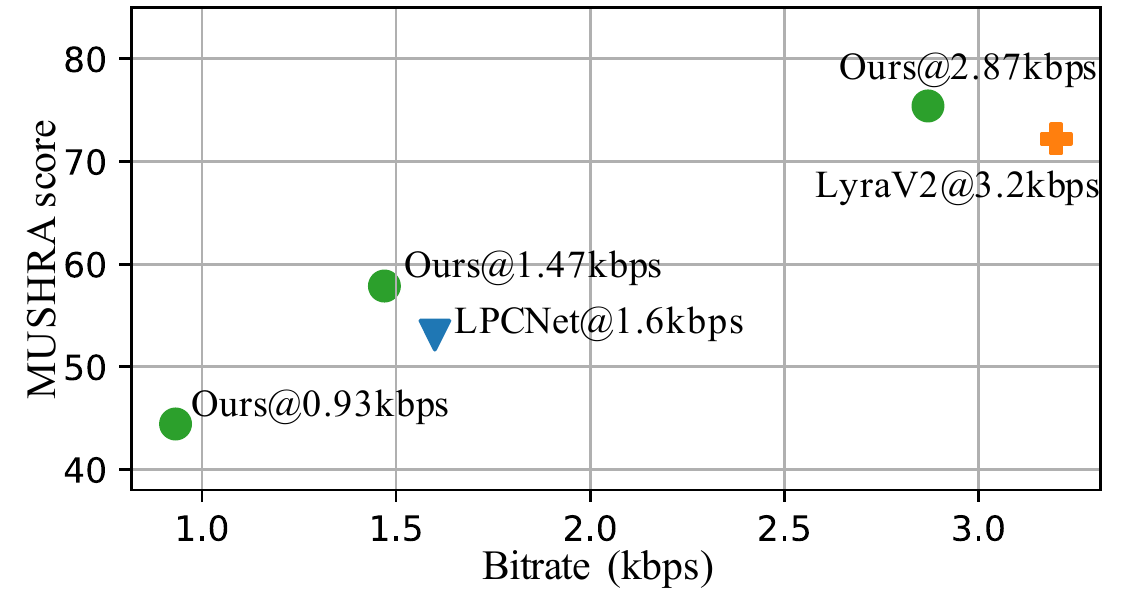}
        \vspace{-4mm}
        \caption{MUSHRA score. The reference at MUSHRA score 100 is not shown in the graph.}
        \label{fig:subjective}
        \vspace{-4mm}
\end{figure}

\begin{figure}[t]
        \centering
        \includegraphics[width=\columnwidth]{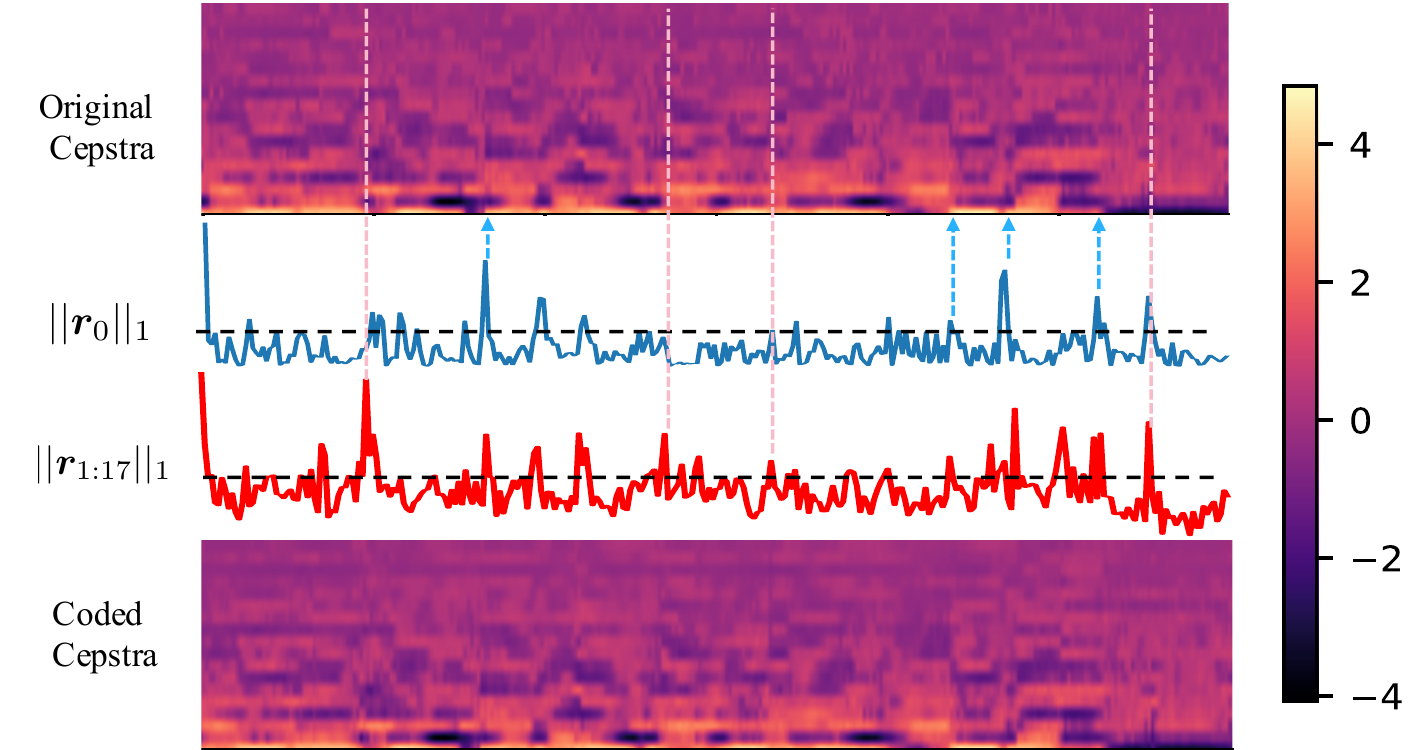}
        \vspace{-2mm}
        \caption{The original cepstra, coded cepstra, the $L1$ norm curve for the vector residuals (red), and the $L_1$ norm curve for the scalar residuals (blue) of a 3s sample. The black dash lines mark the thresholds for the 1.48 kbps case. }
        \label{fig:cepstral}
        \vspace{-6mm}
\end{figure}

\subsection{Subjective test results}
Fig. \ref{fig:subjective} shows the scores from the MUSHRA test. We can see that the proposed model outperforms LPCNet at a lower bitrate ($\sim$ 1.47 vs. 1.6 kbps). At $\sim$ 0.93kbps, our codec is slightly worse than LPCNet but its performance could be acceptable given the 40\% of bitrate reduction. The proposed codec at 2.87 kbps is perceptually better than Lyra V2 at $3.2$kbps. The results demonstrate the proposed model’s scalability and effectiveness across different bitrate ranges. 

\subsection{Analysis on the discriminative residual coding}
In Fig.~\ref{fig:cepstral}, we picked a random utterance sample and aligned its cepstra with the coded version of the cepstra. In between, it lays the $L_1$ norm curves of the scalar and vector residuals in blue and red. At the points where any curve is over the black dash threshold, the predictor failed to make a good prediction, thus requiring more bits to represent these residuals using a $\mathcal{Q}_\text{L}$ scheme. Those properly coded frames take up only $7 \%$ of the total frames in this $\sim 1.48$ kbps case. As for the below-threshold area, conversely, the quantization falls back to the $\mathcal{Q}_\text{S}$ mode while being frugal in assigning bits to these well-predictable frames.

We made three main observations from the graph. Firstly, our codec does a good job estimating and coding the original cepstra (by comparing the top and bottom cepstra), especially at the lower dimensions, although with some energy loss at the higher dimension. Even at the low-bitrate coding area, the cepstral patterns are still well captured. 
Secondly, we can observe that the curves' peaks align with the original cepstrum's transient events. To show the matches, we marked some arrows and dash lines that connect the residual $L_1$ norm curves to the cepstra as examples. The alignment of cepstral changes and the peaks indicates that, although the performance of the feature predictor degrades at the transient events, the discriminative coding can make accurate compensation at a minimal cost.
Finally, it is also noteworthy that despite embracing similarity, the $L_1$ norm of the vector and scalar can have peaks and valleys at different places. Hence, they could compensate for the predictor model's different behavior at individual subbands, which also justifies the advantage of band-specific discriminative coding for the SQ and VQ parts.

%% file: conclusion.tex
\section{Conclusion}

In this work, we proposed a lightweight, low-latency, low-bitrate speech coding framework. In line with the parametric coding paradigm, we designed a feature predictor to capture the temporal redundancy and reduce the burden of coding raw feature frames. Moreover, we applied the discriminative coding scheme to the residual signal to further improve coding gain.
We showed that the proposed combination of predictive coding and discriminative residual coding can be harmonized well with the original LPCNet-based codec by providing a more effective quantization scheme than the original multi-stage VQ. We open-source our codes at \url{https://saige.sice.indiana.edu/research-projects/predictive-LPCNet}. 